\documentclass[4paper]{article}
\pdfoutput=1
\usepackage{jheppubmod}
\usepackage{epsfig,amsmath,latexsym,amssymb}

\newcommand{\be}{\begin{equation}}
\newcommand{\ee}{\end{equation}}
\newcommand{\bea}{\begin{eqnarray}}
\newcommand{\ena}{\end{eqnarray}}
\newcommand{\no}{\noindent}
\newcommand{\nb}{\nonumber}

\renewcommand\o{\omega}

\newcommand\m{\ensuremath{\mu}}

\newcommand\n{\ensuremath{\nu}}

\newcommand\tr{\text{tr}}

\newcommand{\de}{\partial}

\newcommand{\ba}{\begin{eqnarray}}
\newcommand{\ea}{\end{eqnarray}}

\title{Perturbations in Massive Gravity Cosmology} 
\author[a]{D. Comelli, }
\author[b,c]{ M. Crisostomi, }
\author[b,c]{and L. Pilo}
\affiliation[a]{ INFN, Sezione di Ferrara,  I-35131 Ferrara, Italy}
\affiliation[b]{Dipartimento di Fisica, Universit\`a di L'Aquila,  I-67010 L'Aquila, Italy}
\affiliation[c]{INFN, Laboratori Nazionali del Gran Sasso, I-67010
  Assergi, Italy}
\emailAdd{comelli@fe.infn.it}
\emailAdd{marco.crisostomi@aquila.infn.it}
\emailAdd{luigi.pilo@aquila.infn.it}

\date{\small \today}

\abstract{
We study cosmological perturbations for  a ghost free massive gravity
theory formulated with a dynamical extra metric that is needed to massive deform GR.
In this  formulation  FRW background solutions fall in
two branches.
In the dynamics of perturbations around the first branch solutions, no extra degree of freedom with respect to GR is
present  at linearized level, likewise what is found in the
Stuckelberg formulation of massive gravity where the extra metric is
flat and non dynamical. In the first branch, perturbations are probably strongly coupled.
On the contrary, for perturbations around the second
branch solutions all expected degrees of freedom  propagate.
While  tensor and vector perturbations of the physical metric that
couples with matter follow closely the ones of GR,  
scalars develop an exponential  Jeans-like instability on sub-horizon scales.
On the other hand, around a de Sitter  background there is no instability.
We argue that one could get rid of the instabilities by
introducing a mirror dark matter sector minimally coupled to only the
second metric.} 

\begin{document}
\maketitle
\section{Introduction}
Recently, there has been a renewed interest in the search of a
modified theory of gravity at large distances through a massive
deformation of GR  (see for a recent review ~\cite{Hint}). A great
deal of effort was devoted  to extend at the
nonlinear level~\cite{Gabadadze:2010} the seminal work of 
Fierz and Pauli (FP)~\cite{Fierz:1939ix}.  The FP theory is defined at linearized level and is
plagued by a number of diseases. In particular, the modification of
the Newtonian potential is not continuous when the mass $m^2$
vanishes, giving a large correction (25\%) to the light deflection
from the sun that is experimentally excluded~\cite{DIS}. A possible
way to circumvent the the discontinuity problem is to suppose
that~\cite{Vainshtein} the linearized
approximation breaks down near a massive object like the sun and an improved perturbative
expansion.
In addition, FP is problematic as an effective theory. Regarding FP as
a gauge theory where the gauge symmetry is broken by a explicit mass
term $m$, one would expect a cutoff $\Lambda_2 \sim m g^{-1} = (m
M_{pl})^{1/2}$, however the real cutoff is $\Lambda_5 = (m^4
M_{pl})^{1/5}$ or $\Lambda_3 = (m^2 M_{pl})^{1/3}$, much
lower than $\Lambda_2$ \cite{AGS}. A would-be Goldstone mode is responsible for
the extreme $UV$ sensitivity of the FP theory, that becomes totally
unreliable in the absence of proper UV completion.  
Recently it was shown that there exists a non linear completion of the FP
theory~\cite{Gabadadze:2011} that is free of ghosts up to the fourth
order \cite{Gabadadze:2011}, avoiding the presence of the Boulware-Deser
instability~\cite{BD}.  Then the propagation of only five degrees of freedom (DoF)
was generalized to all orders in~\cite{GF}; this was shown also in the Stuckelberg language
in~\cite{deRham:2011qq}. 

Quite naturally massive gravity leads to bigravity.  Indeed, any
massive deformation, obtained by adding to the Einstein-Hilbert
action a non-derivative self-coupling for the metric $g$, requires the
introduction of an additional metric $\tilde g$ that may be a fixed
external field, or be a dynamical one.  
When $\tilde g$ is non-dynamical we are in the framework  of 
\ae ther-like theories; on the
other hand if it is dynamical we enter in the realm of
bigravity~\cite{DAM1} that was originally introduced by Isham, Salam
and Strathdee~\cite{Isham}.  The need for a second
dynamical metric also follows from rather general grounds. Indeed, it
was shown in~\cite{DeJac} that in the case of non singular static
spherically symmetric geometry with the additional property that 
the two metrics are diagonal in the same coordinate patch, a Killing
horizon for $g$ must also be a Killing horizon for $\tilde
g$; see~\cite{gabBH} for a concrete example. Actually it turns out that the off diagonal solutions show no
modification of gravity at large distance~\cite{us,ussphe}.
Also cosmology calls for the bigravity formulation of massive
gravity. When the second metric is static 
there is no homogeneous spatially flat FRW solution~\cite{FRWfroz,Alberte,cur}, 
on the contrary  in the bigravity formulation flat FRW homogeneous solutions do
exist~\cite{uscosm,hasscosm,russ}. See also~\cite{Koy-cosm} for a
different approach to cosmology in massive gravity.
In this paper we study perturbations around the FRW background
solutions found in~\cite{uscosm}.

The outline of the paper is the following.
After a brief introduction to the bigravity formulation of massive
gravity in section~\ref{bi}, the FRW solutions are reviewed in section~\ref{frwsec}. 
The perturbed FRW geometry is introduced in section~\ref{pert}  and the perturbed Einstein equations are given in
section~\ref{modeq}.  The dynamics of the perturbations in various
cases are studied in sections \ref{br1}, \ref{ds} and 
\ref{br2}. Section~\ref{con} contains our conclusions.

\section{Massive Gravity and Bigravity}
\label{bi}
Any modification of GR that turns a massless graviton into a massive
one calls for additional DoF. An elegant way to provide them is to
work with an extra tensor $\tilde g_{\mu \nu}$. When coupled to the
standard metric $g_{\mu \nu}$, it allows to build non-trivial
diff-invariant operators that lead to mass terms  when expanded around
a background.  Consider the action
\be
S=
\int d^4 x \left\{ \sqrt{\tilde g} \,   \kappa   \, M_{pl}^2\; \tilde {\cal R}+\sqrt{g} \;\left[ 
M_{pl}^2 \;\left( {\cal R}
-2 \, m^2   \, V \right)  + L_{\text{matt}} \right] \right\},
\label{act} 
\ee
where $R(g_i)$ are the corresponding Ricci scalars and the interaction
potential $V$ is a scalar function of the tensor $X^\mu_\nu = {g}^{\mu
\alpha} {\tilde g}_{\alpha \nu}$.  Matter is minimally coupled to
$g$ and it is described by $L_{matt}$. The constant $\kappa$ controls
the relative size of the strength of gravitational interactions in the
two sectors, while $m$ sets the scale of the graviton mass. The action
(\ref{act}) brings us into the realm of bigravity theories, whose
study started in the '60 (see \cite{DAM1} for early references).  An
action of the form (\ref{act}) can be also viewed as the effective
theories for the low lying Kaluza-Klein modes in brane world
models~\cite{DAM1}. The massive
deformation is encoded in the non derivative coupling between
$g_{\mu\nu}$ and the extra tensor field $\tilde g_{\mu \nu}$. Clearly
the action is invariant under diffeomorphisms, which transform the two
fields in the same way (diagonal diffs).
Taking the limit $\kappa \to \infty$, the
second metric decouples, and gets effectively frozen to a fixed
background value so that the ``relative'' diffeomorphisms are
effectively broken, as far as the first metric is concerned.
Depending on the background value of $\tilde g_{\mu \nu}$ one can
explore both the Lorentz-invariant (LI) and the Lorentz-breaking (LB) phases of
massive gravity~\cite{PRLus} \cite{blasmy}. When the second metric is dynamical this is
determined by its asymptotic properties, as discussed below. In this
case notice that $\tilde g_{\mu \nu}$ is determined by its equations
of motion (for any finite $\tilde M_{pl}$) so that we will be working
always with consistent and dynamically determined backgrounds.  The
role played by $\tilde g_{\mu \nu}$ is very similar to the Higgs
field, its dynamical part restores gauge invariance and its background
value determines the realization of the residual symmetries.

The modified Einstein equations can be written as\footnote{When not
specified, indices of tensors related with $g$($\tilde g$) are raised/lowered with
$g(\tilde g)$}
\begin{gather}
\label{eqm1}
\,{E}^\mu_\nu  +  Q_1{}^\m_\n = \frac{1}{2\;M_{pl}^2 }\, {T}^\mu_\nu   \\
\label{eqm2}
\kappa  \, {\tilde E}^\mu_\nu  +  Q_2{}^\m_\n = 0 \; ;
\end{gather}
where we have defined $Q_1$ and $Q_2$ as effective energy-momentum tensors induced by the
interaction term. The only invariant tensor that can be written without derivatives out
of $g$ and $\tilde g$ is $X^\mu_\nu = g^{\mu \alpha}_1 {\tilde g}_{\alpha
\nu}$ \cite{DAM1}. The ghost free
potential~\cite{Gabadadze:2011}\footnote{A very similar potential
  having the same form but with $X$ instead of $X^{1/2}$ was
  considered in~\cite{us}.}  $V$ is  a special  scalar function 
of $Y^\mu_\nu=(\sqrt{X})^\mu_\nu$ given by
\be
\label{eq:genpot}
\qquad V=\sum_{n=0}^4 \, a_n\, V_n \,,\qquad n=0\ldots4 \, ,
\ee 
where The $V_n$ are the symmetric polynomials of $Y$
%
%\vspace*{-1ex}
\be
\begin{split}
&V_0=1\,\qquad 
V_1=\tau_1\,,\qquad
V_2=\tau_1^2-\tau_2\,,\qquad
V_3=\tau_1^3-3\,\tau_1\,\tau_2+2\,\tau_3\,,\\[1ex]
&V_4=\tau_1^4-6\,\tau_1^2\,\tau_2+8\,\tau_1\,\tau_3+3\,\tau_2^2-6\, \tau_4\,,
\end{split}
\ee 
with $\tau_n=\tr(Y^n)$.  In \cite{HRBI} it was shown that in the bimetric
formulation the potential $V$ is ghost free. We have that 
\bea
\label{eq:q1}
 Q_1{}_\nu^\mu &=&  { m^2}\, \left[ \;  V\; \delta^\mu_\nu \,  - \,  (V'\;Y)^\mu_\nu  \right]\\[1ex]
\label{eq:q2}
 Q_2{}_\nu^\mu &=&  m^2\, q^{-1/2} \, \; (V'\;Y)^\mu_\nu ,
\ena
where  $(V^\prime)^\mu_\nu = \de V / \de Y_\mu^\nu$ and  $q =\det
X=\det(\tilde g)/\det(g)$.  

The canonical analysis~\cite{GF} shows that in general 7 DoF propagate; around a
Minkowski background, 5 can be associated to a massive spin two
graviton and the remaining 2 to a massless spin two graviton.

\section{FRW Solutions in Massive Gravity}
\label{frwsec}
Let us review the FRW background solutions in massive gravity~\cite{uscosm} that are of the form 
\be
\begin{split}
ds^2 &=   a^2(\tau) \left(- d \tau^2 +   dr^2 + r^2 \, d \Omega^2 \right) =
 {\bar g}_{1 \, \mu \nu} dx^\mu dx^\nu\\
\tilde{ds}^2 &= \omega^2(\tau) \left[- c^2(\tau) \, d \tau^2 + dr^2+  r^2 \, d
  \Omega^2 \right] =
 {\bar g}_{2 \, \mu \nu} dx^\mu dx^\nu\, .
\label{frw}
\end{split}
\ee
It is convenient to define the  standard Hubble 
parameters for the two metrics
\be
{\cal H} =\frac{d a}{d\tau}  \frac{1}{a} \equiv \frac{a'}{a}= H \, a  \, ,
\qquad {\cal H}_\o = \frac{\o'}{\o} = H_\o \, \o \, , \qquad \xi =
\frac{\o}{a} \, .
\ee
Solutions fall in two branches depending on how the Bianchi
identities are realized. 
%\vspace{0.5cm}
\begin{itemize}
\item
In branch one, $\xi=\bar \xi$ is constant and satisfies the following algebraic equation 
\be 
f_2(\bar \xi) =0 \, \qquad \text{with } f_2(\xi)= 6\, a_3 \, \xi^2+4\, a_2 \, \xi +a_1  .
\label{cb1}
\ee
As a consequence, the Hubble parameter ${\cal H}$ of $g$ and
the one of $\tilde g$, ${\cal H}_\o$ coincide and
\be
\frac{{\cal H}^2}{a^2} =\frac{8 \pi G}{3}  \left( \rho  +
  \Lambda_{1} \right) \, , \qquad \Lambda_{1} =\frac{ m^2 }{ 8 \pi G}  \left[ a_0-6 \,{\bar \xi}^2\,
  \left(2\, a_3 \,\bar \xi +a_2 \,\right)\right] \; .
\label{tteq}
\ee
In this branch the effect of the mass deformation is to induce an
effective cosmological constant and 
\be
c =\sqrt{\frac{\Lambda_{1}+\rho }{\Lambda_2}}\qquad {\rm with}\quad  \Lambda_{2} =\frac{ m^2 }{ 4 \pi G\,\kappa}   \left[6\, \bar \xi\,  \left(2\, a_4 \,\bar \xi
   +a_3 \, \right)+a_2\right]  \, .
\ee
This is not very surprising,
indeed, the constraint (\ref{cb1}),  in the spherically symmetric case,
leads  to a branch of solutions with no modification of gravity,
being the graviton mass zero around a flat background~\cite{ussphe}. 
\item
In branch two, $\xi$ is not constant and  the Bianchi identities are realized in the form
\be
c = \frac{H_\o}{H} \, \xi =  \frac{{\cal H}_\o}{{\cal H}} ,\;\;\;\;\;\; \xi'=(c-1)\;{\cal H}\;\xi 
\;\;\;\;  {\rm with}\;\;\; c>0 \, , 
\label{bnew}
\ee 
and 
\be
\frac{3 \mathcal{H}^2}{a^2}= 8 \pi  G \, \rho  + m^2 \left(6 a_3 \,
  \xi ^3+6 a_2 \,  \xi^2+3 a_1 \,  \xi +a_0 \right) \, .
\ee
The ratio $\xi$ of the two scale parameters satisfies the equation
\ba%\nonumber
%\begin{split}
 &&m^2 \left[\xi ^2 \left(\frac{8\; a_4}{\kappa }-2\; a_2\right)+\xi 
   \left(\frac{6 \; a_3}{\kappa }-a_1\right)+\frac{a_1}{3 \; \kappa \, 
   \xi }+\frac{2 \; a_2}{\kappa }-2\;  a_3 \, \xi ^3-\frac{a_0}{3}\right] 
   %\\&&
=\frac{8 \pi  G \; \rho  }{3}  \, .
%\end{split}
\label{conII}
\ea
\end{itemize}
In the expanding universe, the ratio $\frac{8 \pi  G \; \rho  }{m^2}$
scales rapidly with redshift $z$, $1+z\equiv
a(\tau_{today})/a(\tau)$. Indeed, 
assuming that the mass scale $m$ is related to the present
cosmological constant as $\; m^2\,M_{pl}^2\propto \Lambda$ and that
for matter with an equation of state $p  = w \, \rho $,  $\rho  =
\rho_0 \, (1+z)^{3(w+1)}$, we have 
\be
\frac{8 \pi  G \; \rho  }{m^2}\sim \frac{\Omega_m}{\Omega_{\Lambda}}
\;z^{3(w+1)}\gg 1 \quad {\rm at\; early\; times.}
\ee
This quantity gets very large already at redshift $z\sim 10$,
much later than the radiation era ($z\sim 10^4$).
Thus, eq.(\ref{conII}) can be satisfied only in two regimes: 
for  large or   small values of $\xi$~\footnote{In presence of
  ``mirror'' matter with energy density $\tilde \rho$ minimally
  coupled with the  second metric $\tilde g$,  eq.(\ref{conII}) gets
  modified as $\rho \to \rho  -\xi^2\,\tilde\rho /\kappa$ and 
the small-large $\xi$ regime solutions can be evaded as soon as $\xi^2\sim \kappa\,\rho/ \tilde \rho$, when $m \to 0$.}.
The large $\xi$ regime is physically uninteresting because $c<0$ when
matter has $w>0$ \cite{uscosm}.  
In the small   $\xi$  regime,  cosmology is very similar to the standard one;
for  $a_1> 0$ we have
\ba
\xi &=&\frac{a_1\; m^2}{8\,\pi\, G \, \kappa\,\rho } + {\cal O}\left({\frac{m^2}{G\,\rho }}\right)^2 \sim z^{-3(w+1)}\; \, ;\\
&&\rho  +\rho_g =  \rho \left(1+{\cal O}\left(\frac{m^2}{G\,\rho }\right)\right) \, ;\\ 
 && w_{eff} = w +{\cal O}\left(\frac{m^2}{G\,\rho }\right) \, ,
 \quad \qquad c=(4+3\,w)+{\cal O}\left(\frac{m^2}{G\,\rho }\right) \, . 
\ea
Once the matter is so diluted that $\rho $ is negligible in
(\ref{conII}) the system falls in the fixed point region and $\xi$ is
almost constant and the universe  enters in a late time
dS phase. 
%\end{itemize}
The analysis  is identical when the {\it same}  spatial
curvature $k_c$ is introduced in (\ref{frw})\footnote{The spatial
curvatures must be equal for consistency~\cite{uscosm}.} for both metrics.

Let us now discuss the 
differences with the frozen metric approach where
% different approach according with the second metric
 $\tilde g$ is non dynamical.  Formally, the
non-dynamical limit corresponds to $\kappa \to \infty$.
To make contact with the existing literature, we take $\tilde g$ equivalent
to the Minkowski flat metric implying clearly that $c$ and $\omega$
cannot be arbitrary. Indeed, imposing that the Riemann curvature
tensor of   $\tilde g$ vanishes we get that $k_c <0 $ and
\be
c= \frac{{\cal H}_\o}{\sqrt{-k_c}} \, .    
\label{cflat}
\ee
Thus, flat  FRW cosmology with frozen second metric exists only with a negative
non-vanishing spatial curvature~\cite{FRWfroz}-\cite{uscosm}. When
(\ref{cflat}) holds, Bianchi identities can be realized only within 
branch one, leading to eq.(\ref{cb1}).
 That is why FRW perturbations in the Stuckelberg formalism are stuck  into  branch one that is rather problematic. 

\section{Perturbed FRW Geometry}
\label{pert}
Let us now consider the perturbations of the FRW background (\ref{frwsec})
\be
g_{\mu \nu} = \,  \bar g_{1 \, \mu \nu}  +a^2 \,  {h_1}_{\mu
  \nu} \, , \qquad {\tilde g}_{\mu \nu} =  \bar g_{2 \, \mu \nu}  +
 \o^2 \,  h_{2 \, \mu \nu}  \, .
\ee
parametrized as follows 
\be
\begin{split}
& {h}_{1 \, 00} \equiv - 2 A_1  \, , \qquad  {h}_{2 \, 00}  \equiv - 2 c^2 \, A_2\\
& {h}_{1/2 \, 0 i}  \equiv {\cal C}_{1/2 \,i} - \de_i B_{1/2}  \, , \qquad
\de^i {\cal V}_{1/2 \, i}=  \de^i {\cal C}_{1/2 \, i} = \de^j {h^{TT}}_{1/2 \, ij} =
\delta^{ij} {h^{TT}}_{1/2 \, ij}  =0  \, ,\\
& h_{1/2 \, ij}  \equiv {h^{TT}}_{1/2 \, ij} + \de_i {\cal V}_{1/2 \, j} + \de_j {\cal
  V}_{1/2\, i} + 2 \de_i \de_j E_{1/2} + 2 \, \delta_{ij} \, F_{1/2}  \,.
\end{split}
\ee
Spatial indices are raised/lowered using the spatial flat metric.

Under a gauge transformation generated by $\zeta^\mu$ the metric perturbation transforms
\be
\begin{split}
& \delta h_{1 \, \mu \nu } = a^{-2} \left(\zeta^\alpha \de_\alpha  {\bar g}_{1 \,
  \mu \nu} +  \bar g_{1 \, \alpha \nu}  \, \de_\mu
  \zeta^\alpha + \bar g_{1 \, \mu \alpha}  \, \de_\nu \zeta^\alpha \right)
\, , \\[.2cm]
&\delta h_{2 \, \mu \nu } = \o^{-2} \left(\zeta^\alpha \de_\alpha \, {\bar
  g}_{2 \,\mu \nu } +  \bar g_{2 \, \alpha \nu}  \, \de_\mu
  \zeta^\alpha + \bar g_{2 \, \mu \alpha}  \, \de_\nu \zeta^\alpha \right) \, .
\end{split}
\ee
and for the corresponding components 
\be
\begin{split}
&\delta A_1 = {\cal H} \, \zeta^0 + {\zeta^0}^\prime \, , 
\quad \delta B_1 =
\zeta^0 -\zeta^\prime \, , \quad \delta E_1 =  \zeta \, ,  \quad
\delta F_1 =  {\cal H} \, \zeta^0
\, ; \\
&\delta A_2 = {\cal H}_\beta \, \zeta^0 + {\zeta^0}^\prime \, , 
\quad \delta B_2 =
 c^2 \, \zeta^0 - \zeta^\prime \, , \quad \delta E_2 =  \zeta \, ,  \quad \delta F_2 =  {\cal H}_\o \zeta^0
\, ; \\
& \delta {\cal C}_{1/2 \, i} = {\zeta^i_T}' \, , \quad \delta {\cal
  V}_{1/2 \, i} =  {\zeta^i_T} \,  , \qquad \delta {h^{TT}}_{1/2 \, ij}  =0 \, ; \\
\end{split} 
\label{gtr}
\ee 
where 
\be
\begin{split}
&\zeta^i = \zeta^i_T + \de_i
\zeta \, , \qquad\zeta = \Delta^{-1} \de_i \zeta^i \, ,  \\[.2cm]
& {\cal H}_\beta=\frac{ (c
  \, \omega)'}{(c \, \o)} =\frac{c'}{c} +\o \, {\cal H}_\o \, .
\end{split}
\ee
In the scalar sector we have 8 fields and two independent gauge
transformations, as a result we can form 6 independent gauge invariant
scalar combinations  that we chose to be 
\be
\begin{split}
& \Psi_1= A_1 - {\cal H} \, \Xi_1   - \Xi_1^\prime \qquad \Psi_2= A_2 + c^{-2} \left(\frac{c'}{c} - {\cal H}_\o \right) \, \Xi_2   -
\frac{\Xi_2^\prime}{c^2}  \\
&\Phi_1 = F_1- {\cal H} \, \Xi_1 \, , \qquad \Phi_2 = F_2 - {\cal
  H}_\o \, \frac{\Xi_2}{c^2}  \, , \\
&{\cal E} = E_1 - E_2 \, ,  \qquad {\cal B}_1 = B_2 - c^2
B_1 +(1-c^2) \, E_1' \, ,
\end{split}
\label{sgib}
\ee
where $\Xi_{1/2} = B_{1/2} + E_{1/2}^\prime$. The following additional
gauge invariant fields will be useful to write in a compact form the
perturbed Einstein equations
\be
\begin{split}
&{\cal F}_1=F_2-F_1 +\left({\cal H} - {\cal  H}_\o \right) \Xi_1 \, ,
\qquad {\cal F}_2=F_2-F_1 +\left({\cal H} -
{\cal  H}_\o \right) \frac{\Xi_2}{c^2} \, , \\
 &{\cal B}_2 = B_2 - c^2 B_1 +(1-c^2) \, E_2' \, , \\
&{\cal A}_1 = c (A_2-A_1) + \left[c \left({\cal H} -
{\cal  H}_\o \right) - c'\right] \Xi_1 \, , \\
& {\cal A}_2 = c
(A_2-A_1) + \left[c \left({\cal H} -
{\cal  H}_\o \right) - c'\right] \frac{\Xi_2}{c^2} \, .
\end{split}
\ee
The fields ${\cal F}_{1/2}$, ${\cal A}_{1/2}$ and ${\cal B}_2$ can be
expressed in terms of the ones in (\ref{sgib}), as it is shown in 
Appendix~\ref{rel}.  In the matter sector, we define the following gauge
invariant perturbed pressure and density
\be
\delta \rho_{gi} = \delta \rho - \Xi_1 \, \rho' \, , \qquad \delta
p_{gi} = \delta p - \Xi_1 \, p'
 \, .
\ee
For matter, together with pressure and density perturbation, there is
also the perturbed 4-velocity $u^\mu$ that consists of a scalar part
$v$ and a vector part $\delta z_i$
\be
\begin{split}
&u^\mu = {\bar u}^\mu  + \delta u^\mu \, , \qquad u^\mu u^\nu g_{\mu
  \nu} = -1 \, , \quad \delta u^0 = - a^{-1} \,  A_1 \, ;  \\
& \delta u_i = a \left( \de_i v - \de_i B_1 +  \delta z_i + {\cal
    C}_{1 i} \right) \, .
\end{split}
\ee
The corresponding gauge invariant quantity are defined as 
\be
u_s = v + E_1' \, , \qquad {\delta v}_{ i} = \delta z_i + {\cal
  C}_{1 \, i} \, .  
\ee
The conservation of the matter EMT  leads to a set of differential
relations; for scalar matter perturbations we have
\bea
\delta \rho_{gi}'=(1+w) \left[\rho \,\left( k^2 \, u_s-3 \, \Phi_1'
  \right) - 3  \, {\cal H } \, \delta \rho \right] \, ;\\
u_s'= (3 w-1) \, u_s \, {\cal H} - \frac{w}{(1+w)} \, \frac{\delta
\rho_{gi}}{\rho} - \Psi_ 1\, ;
\ena
while for vector matter perturbations 
\be
\delta v_i ' =\delta v_i \, (3 \,w-1)\,
   \mathcal{H} \, .
\label{vecc}
\ee
In the vector sector we have 4 fields and 1 gauge transformation; thus,
we can form 3 independent gauge invariant vector perturbations
\be
\begin{split}
& V_{1/2 \, i} = {\cal C}_{1/2 \, i} - {\cal V}_{1/2 \, i}^\prime \, ,
\quad \chi_i = {\cal C}_{1 \, i}-  {\cal C}_{2 \, i} \, .
\end{split}
\ee
\section{Perturbed Einstein Equations}
\label{modeq}
Let us start with the scalar sector.  The  leading order  perturbed Einstein equations for $g$ are
\bea
&&2 \Delta  \Phi_1+6 {\cal H} \left(\Psi _1 \mathcal{H}-\Phi
  _1'\right) +a^2 m^2 \,  f_2 (3 {\cal F}_1-\Delta   {\cal
  E})  =-8 \pi  a^2 \,G \,\delta \rho_{\text{gi}} \, ; \label{1tt}\\[.3cm]
&&\de_i \left[2 \Psi _1 \mathcal{H}-2 \Phi _1' +\frac{a^2 m^2  \,
    \mathcal{B}_1 \, f_2 }{(c+1)} +  8 \pi\,  G \,a^2\, (p+\rho )\,  u_s\right]
=0 \, ;  \label{1ts}\\[.3cm] 
&&  \left(\de_i \de_j - \delta_{ij} \Delta \right)
   \left( a^2 \,f_1\, m^2 \mathcal{E} - \Phi_1- \Psi_1\right)  + \delta_{ij}
   \left[m^2   \, a^2 \,(2\, f_1\, \mathcal{F}_1  +f_2\, \mathcal{A}_1 )+2
     \Psi_1 \,\left(\mathcal{H}^2+2 \,\mathcal{H}'\right) \right. \nb \label{1ss}\\[.2cm]
&&\left. -2 \,\Phi_1''-2\, \mathcal{H} \left(2\, \Phi _1'-\Psi _1'\right)\right]
= 8 \pi  G \, a^2 \, \delta_{ij} \,  
\delta p_{\text{gi}} \, , 
\ena
where
\be
f_1 =\xi \, \left[2 \,\xi  \,\left(3\, a_3 \, c \, \xi +a_2
 \,  (c+1)\right)+a_1\right] \, , \qquad f_2 =\xi \, \left(6 \,a_3 \,\xi ^2+4
\, a_2\, \xi +a_1\right) \, . 
\ee
For the metric $\tilde g$ we have 
\bea
&&
2\,c^2 \Delta  \Phi _2+6
   \mathcal{H}_{\omega } \left(\Psi _2   \mathcal{H}_{\omega }-\Phi _2'\right)  +\frac{m^2  a^2  f_2}{\kappa\,\xi^2}\, c^2 
\,   \left(\Delta  \mathcal{E}-3 \,\mathcal{F}_2\right)=0 \, ; \label{2tt}\\[.3cm]
&& \de_i \left[ 2\, c \, 
   \left(\Psi _2 \mathcal{H}_{\omega }-\Phi _2'\right)-\frac{m^2  \, a^2 \, f_2}{\kappa\,\xi^2\,(1+c)} \, \mathcal{B}_2\right] =0 \, ;
 \label{2ts}\\[.3cm]
&&
-c \, \left(\de_i \de_j - \delta_{ij} \Delta \right) \left[ 
\frac{a^2 \, f_1 \, m^2}{\kappa\,\xi^2} \, \mathcal{E}+ \, c   \, \left(\Phi_2+\Psi
     _2\right)\right]+\delta_{ij}\left[ \frac{m^2 \,a^2}{\kappa\,\xi^2} (2\,  c\, f_1 \, \mathcal{F}_2+  f_2 \, \mathcal{A}_2 ) +\right.
\nb\\[.2cm]
&&
\left.
2\left( \mathcal{H}_{\omega }^2+2\,\mathcal{H}_{\omega }'-2\,\frac{c'}{c}\, \mathcal{H}_{\omega }\right) \Psi _2-
2 \Phi_2''+2\left(\frac{c'}{c}-2\,\mathcal{H}_{\omega }\right)\, \Phi
_2'+2\,\mathcal{H}_{\omega }\,\Psi _2' \right]=0 \, . \label{2ss}
\ena
For the vector sector the perturbed Einstein equations are
\bea
&& \frac{\Delta V_{1 \, i}}{2 \, a^2} - 8 \pi G  \, (\rho + p)  \, \delta v_i - 
\frac{m^2 }{(1+c)} \, f_2 \, \chi_i =0 \, ; \label{1vts}\\[.3cm]
&&\de_{(i} V_{1 \, j)}'  + 2 \,{\cal H}\,   \de_{(i} V_{1 \, j)}
=m^2 \, a^2\, f_1\, \de_{(i}
  {\cal V}_{12 \, j)}   \,
; \label{1vss} \\[.3cm]
&& \frac{\Delta V_{2 \, i}}{2 \,a^2 \, c } + \frac{ m^2 \, f_2 }{(1+c)\,\kappa\, \xi^2}  \, \chi_i =0 \, ; \label{2vts}\\[.3cm]
&& 
\de_{(i} V_{2 \, j)}' +  \left[2 \left( {\cal H }  +\, \frac{\xi '}{\xi} \right)-\frac{ c'}{ c}\right] \,
  \de_{(i} V_{2 \, j)}   
%\nb \\ && 
+ \, \frac{m^2\,a^2\, c  \, f_1}{\kappa\,\xi^2}\,   \de_{(i}
{\cal V}_{12 \, j)} =0 \label{2vss} \, ;
\ena
where 
\be
{\cal V}_{12 \, i } ={\cal V}_{1\, i} -{\cal V}_{2 \, i } \, , \qquad 
{V}_{12 \, i} ={ V}_{1\, i} -{V}_{2 \, i} \, .
\ee
Notice that ${V}_{12 \,i} = \chi_i -{\cal V}_{12 \, i }' $. 

Finally, for the tensor perturbations we obtain
\bea
&&{h^{TT}}_{1 \, ij}'' + 2 {\cal H} \, {h^{TT}}_{1 \, ij}' - \Delta {h^{TT}}_{1 \, ij} + m^2  \, a^2 \,
f_1 \, \left(h^{TT}_{1 \, ij} -h^{TT}_{2 \, ij} \right) =0 \,
; \label{1t} \\[.3cm]
 && {h^{TT}}_{2 \, ij}'' +\left[2 \left(  {\cal H} +\frac{ \xi'}{\xi} \right)
 -   \,\frac{ c'}{c} \right] {h^{TT}}_{2 \, ij}'  - c^2   \, \Delta {h^{TT}}_{2 \, ij} 
 - \frac{m^2 \,f_1 \, c }{ \kappa\, \xi^2} \, a^2  \,
   \left(h^{TT}_{1 \, ij} -h^{TT}_{2 \, ij} \right) =0 \, .
\label{2t}
\ena
Now that we have at our disposal the full set of equations, 
we can study perturbations around  background solutions in the branch one
and two. In the following we will often use the Fourier transform of
perturbations with the respect to $x^i$, the corresponding
3-momentum will be $k^i$ and $k^2 = k^i k_i$. To keep notation as
simple as possible the symbol of the Fourier transform will be understood.

\section{Branch One Perturbations}
\label{br1}

In this case $\xi=\bar \xi$ is a non-vanishing constant such
that $f_2(\bar \xi)=0$, then ${\cal H}_\o={\cal H}$.  From (\ref{2ts}) we can express $\Psi_2$ in
terms of $\Phi_2$, then from (\ref{2tt}) we get that
$\Psi_2=\Phi_2=0$.
Now, using (\ref{2ss}) we have that ${\cal E}={\cal F}_2=0$ and 
from the relations in Appendix \ref{rel}, also ${\cal F}_1=0$. At this
point it is straightforward to show that using the equations for the scalar
perturbations of $g$ we get 
\be
\left[-w \, \Delta  +(3 w+1) \mathcal{H}^2+2
   \mathcal{H}'\right] \Psi _1 +3 (w+1) \mathcal{H} \Psi _1'+\Psi
 _1''=0 \, 
\ee
Not surprisingly, this equation describes the very same perturbations
of  GR in the  presence of a fluid with an equation of state $w$. 

Also for vectors, being $f_2=0$, again we have the very same equations as in
GR
\bea
&& \Delta V_{1 \, i} - 16 \pi G \, a^2 \, (\rho + p)  \, \delta v_i  =0 \, ; \\[.2cm]
&&\de_{(i} V_{1 \, j)}'  + 2 {\cal H}   \de_{(i} V_{1 \, j)}
=0 
\ena
with $V_{2\,i}={\cal V}_{12\,i}=0$. Clearly, as in GR no vector propagates. In the tensor sector four modes propagates.
From the canonical analysis, 8 DOF are expected but only $5=1+4$ are
accounted for. Thus, a scalar plus a vector are not
present and are probably strongly coupled, at least around a FRW
background.  Strong coupling was also found in the Stuckelberg
approach~\cite{Mukopert}. This is not very surprising, condition (\ref{cb1}) in
flat space is equivalent to set to zero the graviton mass and, as
a consequence, both spherically symmetric and FRW branch one solutions
show no gravity modification~\cite{ussphe,uscosm}  and the extra DoF are frozen.
Interestingly enough, in the Stuckelberg approach where  the second
metric is non-dynamical, only the branch one is available and
strong coupling is unavoidable, another manifestation  of the rather
constrained nature of a theory with a priori given metric.

\section{Branch Two: Perturbations in dS Phase }
\label{ds}
Before  analyzing  branch two case  in  full generality, it is instructive to consider a
particular limit of it:  a de Sitter  (dS) background, for which we have  
\bea
&&\rho =\text{const.}\;\;  \Rightarrow \;\; \xi=\text{const.}\;\;   
\Rightarrow \;\; c=1 \, ,\;\; f_1=f_2,\;\;\;\; {\cal H}_\o = {\cal
  H}\equiv H\;a \, , \;\; 
%{\rm with}\;\;  H=\frac{m}{\sqrt{3\,\kappa}} \sqrt{12 \xi  \left(2 a_4 \xi +a_3\right)+2
  % a_2+\frac{f_1}{\xi ^2}}.
\\\nonumber
&&{\rm with}\;\;  H=\frac{m}{\sqrt{3\,\kappa}} \sqrt{12\, \xi\,  \left(2\, a_4 \,\xi +a_3\right)+2\,
   a_2+\frac{f_1}{\xi ^2}} \, .
\ena
The dS phase is a fixed point of the FRW geometry of the  branch two solution~\cite{uscosm}.
Introducing 
\be
\begin{split}
&\Phi_1=\frac{1}{2}(\Phi_+ +\Phi_-), \qquad
\Phi_2=\frac{1}{2\,\kappa\,\xi^2}(\Phi_+ -\Phi_- ) \, , \\
&\Psi_1=\frac{1}{2}(\Psi_+ +\Psi_-),\qquad
\Psi_2=\frac{1}{2\,\kappa\,\xi^2}(\Psi_+ - \Psi_-) \, ;
\end{split}
\ee
after some  tedious computations one can show  that all the equations
in the scalar sector are equivalent to a single second order equation  for $\Phi_-$
\bea
%\begin{split}
\Phi _-'' &+&2 \mathcal{H} \, \Phi _-'  \left[\frac{2 k^4}{9 \, a^2 \,
    \mathcal{H}^2 \, 
   m_{\Phi }^2+k^4-18 \mathcal{H}^4}-1\right]+\\ [.2cm]\nonumber
&&\frac{1}{3} \, \Phi_- \left[\frac{4 \left(k^6-3 k^4 \,
      \mathcal{H}^2\right)}{9 \, 
   a^2 \, \mathcal{H}^2 \, m_{\Phi }^2+k^4-18 \mathcal{H}^4}+3 \, a^2 \,
   m_{\Phi }^2-k^2-6 \mathcal{H}^2\right] =0 \; ;
%\end{split}
\label{eqs}
\ena
where
\be
m_{\Phi}^2=m^2  \, f_1 \left(\frac{1}{\kappa\,  \xi
  ^2}+1\right)
\ee
is the mass of the scalar field $\Phi_-$. We have also replaced all
space derivatives $\de_m$ with $i \, k^m$ and $k^2= k^i k^j
\delta_{ij}$.  Defining $\Phi_-= \alpha(t) \, \varphi$, one can choose
$\alpha$ such that the equation for $\varphi$ is canonical
\be
\varphi'' + m^2_\varphi \, \varphi = 0 \, ;
\ee
where
\be
m_\varphi^2 =\frac{4/3 \, 
   k^6+6 k^4 \, \mathcal{H}^2}{9 a^2 \, \mathcal{H}^2 \, m_{\Phi
   }^2+k^4-18 \, \mathcal{H}^4}-\frac{12 k^8 \,
   \mathcal{H}^2}{\left(9 a^2 \, \mathcal{H}^2 \, 
   m_{\Phi }^2+k^4-18 \, \mathcal{H}^4\right){}^2}+a^2 \, m_{\Phi
}^2-\frac{k^2}{3}-2 \, 
   \mathcal{H}^2 \, .
\ee
In the small $k$ limit we have that $m^2_\varphi>0$ when $m^2_\Phi > 2
H^2$ that is precisely the Higuchi bound~\cite{Higuchi:1986py} in dS
spacetime. In the UV (large $k$), $m^2_\varphi>0$ when  $k^2_{\text{ph}}> 8
H^2$; where $ k_{\text{ph}} = a \, k$. In general, one can check that $m^2_\Phi >
2.3353 \,  H^2$ is sufficient to have $m^2_\varphi>0$  for any $k$.

All remaining fields can be written in terms of $\Phi_-$:
\be
\begin{split}
&\Phi_+ = \Psi_+ =0 \, , \qquad \Psi_- =\frac{1}{2\, \mathcal{H}} \left(2
  \Phi _-' -m^2 \,  a^2 \,  f_1 \,  \mathcal{B}_1 \right) \, ;\\[.2cm]
&{\cal B}_1 = \frac{\Phi _- \left(\kappa  \, \xi ^2+1\right) \left(2
\, k^2+3 \, a^2 \, 
\, m_\Phi \right)}{3\,  a^2 \, \, \kappa  \, m_\Phi^2 \, \xi ^2 \, 
\mathcal{H}}-\frac{2 \, k^2 \, \mathcal{E}}{3 \mathcal{H}}+2 \,
\mathcal{E}' \, ; \\[.2cm]
&{\cal E} = \frac{9 \, \mathcal{H} \left(\kappa  \xi ^2+1\right)
  \left(a^2 m_{\Phi }^2-2 \mathcal{H}^2\right) \, \Phi_- '+
  \left(\kappa  \, \xi ^2+1\right)
  \left[ 3 \left(k^2+3 \mathcal{H}^2\right)
   \left(a^2 m_{\Phi }^2-2 \mathcal{H}^2\right)+2
   k^4\right] \, \Phi_- }{2\,  a^2 \, \kappa \, 
   \xi ^2 \, m_{\Phi }^2 \left(9 \, a^2 \, \mathcal{H}^2 \, m_{\Phi
   }^2+k^4-18 \, \mathcal{H}^4\right)} \, .
\end{split}
\ee
As a result, just a single scalar DoF propagates. 

For what concerns the vector sector, all  vectors can  be expressed in terms of ${\cal V}_{12 \, i}$
\be
V_{2 \, i}=-\frac{m_{\Phi}^2 \,  a^2  \, 
   \mathcal{V}_{12 i}'}{ (1+ \xi^2 \, 
   \kappa)  ( a^2  \, m^2_\Phi +  k^2)} \, , \qquad  V_{1 \, i} =- \kappa \, \xi^2 V_{2 \, i} \, ,
\ee
with  ${\cal V}_{12 \, i}$ satisfying a second order equation
\be
\mathcal{V}_{12 i}'' + \frac{2 \, 
   \mathcal{H} \, \mathcal{V}_{12 i}' \left(2  \, 
   k^2+a^2 \, m_\Phi^2  \right)}{  \, k^2+ a^2 \, m_\Phi^2}+
\mathcal{V}_{12 i} \left( k^2+ a^2 \, m_\Phi^2 \right)=0 \, .
\ee
In the tensor sector two modes are propagating (4 DoF). 
The combination $h_{+\,ij}^{TT} = h_{1 \,ij}^{TT} +\xi^2 \, \kappa  \, h_{2 \, ij}^{TT} $ is massless and satisfies
\be
{h_{+ \,ij}^{TT}}''  + 2\,  {\cal H} \, {h_{+ \, ij}^{TT}}' + k^2 \, h_{+ \, ij}^{TT}=0 \, .
\ee
The previous equation is the same for tensor perturbations in GR. While  for the orthogonal combination  $h_{-
  \,ij}^{TT} = h_{1 \,ij}^{TT} - \xi^2 \, \kappa  \, h_{2 \, ij}^{TT}$, we get 
\be
{h_{- \,ij}^{TT}}'' +2 \, \mathcal{H} \, 
   {h_{- \,ij}^{TT}}'+ \left( k^2 + a^2\,  m_{\Phi}^2  \right) \,   h_{- \,ij}^{TT} +
  a^2 \,   m_{\Phi}^2   \frac{\left( \xi^2 \kappa -1\right)}{\left( \xi^2 \kappa +1\right)} h_{+ \,ij}^{TT}=0.
\ee
Summarizing, in the dS phase we have one scalar, one vector and two tensors
that propagate, for  a total of  $1_S+2_V+2_T+ 2_T=7$ DoF,   showing that
all the expected  DoF from the canonical analysis are propagating at
perturbative level. We stress that this is not the case for the branch
one perturbations.
As we will see in the next section, the dynamics of branch two is similar except the
presence of a matter fluid provides an additional scalar DoF.

\section{Branch Two Perturbations}
\label{br2}
The dynamics of branch two perturbations is similar to the dS phase,
only more involved being $\xi$ not constant and $c\neq 1$. 
Due to the complexity of the equations it is difficult and physically
not very interesting to study them for generic values of $\xi$. 
As shown in~\cite{uscosm} and summarized in section~\ref{frwsec},
from early times to redshift of order one, the massive gravity FRW
background solutions are characterized by  a small  value of $\xi$.
As a result, when $\xi <<1$, the background solutions can be expanded in
series of  the  dimensionless ratio
$\tau/\tau_U \ll1$, where  $\tau_U$  is the  age of the universe in
conformal time. For instance, in the radiation dominated era
\be
a = \frac{\tau}{\tau_U} + \epsilon\,\frac{a_0}{10} \,  \left(\frac{\tau}{\tau_U} \right)^5 \, ,
\ee 
where $\epsilon=\frac{m^2}{8 \pi  G \rho
  _0}=\frac{1}{3}\,m^2\,\tau_U^2\;$~\footnote{If we believe that our
  theory is
 origin of  Dark Energy we have to take as reference value 
$m \sim 10^{-33}\,{\rm eV}\sim H_0$ so that 
$\epsilon\sim\frac{\Omega_{\Lambda}}{\Omega_m}$.}.
In Appendix~\ref{corr}, we give the explicit
expressions for the scale factor  $a$ including the leading and   next
to leading terms for both a radiation and matter dominated universe.

An interesting aspect to discuss is the  $m\to 0$ limit.
Naively, taking the formal limit $m \to 0$ in the  equations of motion of
section (\ref{modeq})  we get the perturbed Einstein
equations for two separated GR copies. 
Among the perturbations of metric $g$ that is coupled
  with matter, one scalar (induced by the presence matter)  plus two
transverse and traceless  tensors  propagate. 
In the sector of the perturbations of $\tilde g$ that have no matter sources, only two tensor modes
propagate. However, when we take into account that,
using the background equations, $\xi$ has
a non trivial dependence on $m$, the very same limit is less
straightforward. Indeed, the branch two background solutions are
characterized by a value of  $\xi$ proportional~\footnote{Recall that
at leading order $\xi=\frac{a_1\,m^2}{8\pi G\,\rho}$.} to $m^2$ and this 
completely changes the nature of the  limit $m\to 0$.
  
In the equations for the $g$  perturbations, see  for instance
(\ref{1tt}-\ref{1ss}),  the coupling with the ones of $\tilde g$ is
through the  effective coupling: 
$m^2\,a^2\,f_1$.  
In  the small $\xi$ regime, we have
\be
 m^2\,a^2\,f_1 \approx
a^2\,a_1\,m^2\,\xi  \approx
 a^2\,\frac{a_1^2\,m^4}{8\,\pi\,G\,\rho_m}\,
 \stackrel{m\to 0}{\to}\,0 \, .
\ee
Thus, in the small $m$ limit, 
the perturbations of  the metric $g$   precisely coincide with the
corresponding in GR.
For the equations that govern the perturbation of $\tilde g$, see for
instance (\ref{2tt}-\ref{2ss}),  the effective coupling with $g$ is
$\frac{m^2\,a^2\,f_1}{\kappa\,\xi^2}$.
Then we have that  
\be
\frac{m^2\,a^2\,f_{1/2}}{\kappa\,\xi^2} 
\approx
\, \frac{a^2\,a_1\,m^2}{\,\xi} \,
\approx a^2\,8\,\pi\,G\,\rho_m\,
 \stackrel{m\to 0}{\to}
\,{\rm finite} \, .
\ee
As a result, 
$\tilde g$ perturbations are not GR like in the $m \to 0$ limit,
moreover, as we will show they exhibit a non
trivial structure in the momentum $k$ that is  very different from the 
linear structure in $k^2$ of GR.
This peculiar behaviour  stems  from the interplay of the branch two background
and the structure of perturbed Einstein equations;  
we will return later on this point.

\subsection{Scalar perturbations}
In the scalar sector, all the fields  ${\cal E}, \, {\cal B}_1$
 and $\Psi_{1/2}$ are non dynamical and can   be  expressed in
terms of $\Phi_{1/2}$ that satisfy two second order equations; thus  2 scalar DoF propagate. 
Let us consider the case of a radiation  dominated universe. The
equations of motion for the two propagating scalars have the following structure
\bea
\Phi _a'' &+&\frac{1 }{\tau }\;{\cal D}_{a\,b} \;\;  \Phi _b'+ \frac{1}{\tau^2} \;
{\cal M}_{a\,b}\; \;\Phi _b=0\qquad\qquad  a,\,b=1,2 \, ,
\label{scal}
\ena
where ${\cal D}_{a\,b} $ and ${\cal M}_{a\,b} $ are functions of the
following dimensionless arguments 
$\frac{\tau}{\tau_U}$ , $\epsilon$ and $ x = k \, \tau$. 
%and 
%${\cal M}_{a\,b} ={\cal M}_{a\,b}(\frac{\tau}{\tau_U},\, \epsilon,\,x) $  with $x=k\,\tau$.
% and $\epsilon=\frac{m^2}{8 \pi  G \rho _0}=\frac{1}{3}\,m^2\,\tau_U^2\;$
% \footnote{If we believe that our theory is origin of  Dark Energy we have to take as reference value 
%$m \sim 10^{-33}\,{\rm eV}\sim H_0$ so that 
%$\epsilon\sim\frac{\Omega_{\Lambda}}{\Omega_m}$.}
Note that well in the radiation era (and also matter era ), we have
$\frac{\tau}{\tau_U}\ll 1$; thus, we can expand such a complicated expressions obtaining
at leading order
\bea
\Phi _1'' &+&\frac{4 }{\tau } \;\;  \Phi _1'+ \frac{k^2}{3} \,\; \;\Phi _1+{\cal O}\left(\frac{\tau}{\tau_U} \right)
=0 \, ;\\[0.3cm]\nonumber
\Phi _2'' &+& \frac{10\,x^2+42}{\tau\;(x^2+3)} \;\;  \Phi _2'
+\frac{-5\,x^6-15\,x^4+333\,x^2+999}{3 \, \tau^2\;(x^2+3)^2} \;\;  \Phi _2 -\\&&  \frac{36}{\tau\;(x^2+3)} \;\;  \Phi _1'-
 \frac{3\,(5\,x^2+39) }{\tau^2\;(x^2+3)} \;\;  \Phi _1+{\cal O}\left(\frac{\tau}{\tau_U} \right)
=0 \, .
\ena
The full expressions are lengthy and not particularly illuminating and
are given in  Appendix \ref{ntl} up to the next to the leading order.
We note the various functions ${\cal D}$ and ${\cal M}$ admit a formal expansion in power of $\epsilon$ (i.e $m$)
equivalent on dimensional grounds to an expansion in power of
$\tau/\tau_U$.  
Clearly, the equation  for $\Phi_1$ is the same than GR plus small
corrections, while the equation  for $\Phi_2$ has a non trivial
$k\,\tau$ structure whose origin is the  effective coupling with the
metric $g$ proportional to $m^2/\xi$ that does not vanish in the limit $m\to 0$. In order to get some physical insight, let us consider the case $x<<1$
that physically corresponds to modes well outside the horizon. We have
that 
\bea
\Phi_1 &\sim&  const \, ;\\[0.2cm]
\Phi _2'' &+&\frac{ 14}{\,\tau }\, \Phi _2'+
\frac{  37 \,\, \Phi _2-39 \,\,\Phi_1 }{\,\tau^2 }
=0\;\;\Rightarrow \;\;\Phi_2\sim\frac{39}{37}\, \Phi_1=const \, .
\ena
Thus both scalar perturbations are frozen for the modes well outside
the horizon. On the other hand, in the opposite limit, $x>>1$,
e.g. for the modes well inside the
horizon, we have
\bea
\Phi_1 &\sim& \frac{1}{k^2 \, \tau^{2} }\;\cos k \tau \, ;\\[0.2cm]
\Phi _2'' &+&\frac{ 10  }{\,\tau }\Phi _2'-
\frac{  5\, k^2  }{3 }\Phi _2- 
   \frac{36 }{k^2\,\tau^3 } \Phi _1 '  -\frac{  15 }{\tau ^2  }\Phi
   _1=0 \, , \;\Rightarrow
   \;\Phi_2\sim\frac{1}{(k\tau)^{1/2}}e^{+(\frac{5}{3})^{1/2}
     \,k\,\tau}+ {\cal O}(\Phi_1) \, .
\ena
The solution for the homogeneous  part of the $\Phi_2$ equation has
runaway exponentially behavior. Such a tachyonic instability is due
to the sign of the coefficient of $\Phi_2$, positive for super-horizon
perturbations (and then stable) and negative for sub-horizon  perturbation (and then unstable).
For the matter dominate case (see Appendix \ref{matterpert}) the
situation is very similar  and the same kind of instability for the
sub-horizon modes is present. Notice that such an instability is not present in 
dS case, see eq. (\ref{eqs}).

The leading contribution in the coefficient of $\Phi_2$ for
sub-horizon  modes ( $x >>1$ ) can be computed for a generic equation
of state   $w$, in the small $\xi$ limit. The result is
\be
\left .{\cal M}_{2\,2}\right|_{ x \to \infty}=\left[-  (1+2 \,w)+\frac{2
    \, \left(a_0 \, (w+1) \,\kappa-2 \, a_2\right)}{ \,  a_1}\;\xi
+{\cal O}(\xi^2)\right]\,\, k^2 \, .
\ee
Thus, sub-horizon instabilities are present only when $w>-\frac{1}{2}$
and the $\Phi_2$  perturbation grows exponentially as $e^{(1+2\,w)\,k\,\tau}$.
So, for $w>-\frac{1}{2}$,  the exponential growth of $\Phi_2$ invalidate
perturbation theory at time $\tau \sim 1/k$.  As a consequences,
already in the radiation dominated era sub-horizon perturbations
become  non perturbative, in sharp contrast with GR where matter
perturbations become large only when the universe is non relativistic
due to Jeans instability. 
The other scalar fields are given as a function of $\Phi_{1,\,2}$ in Appendix \ref{scalrad}.

\subsection{Vector perturbations}
In the vector sector, using (\ref{vecc}), as in GR, the velocity
perturbation can be easily obtained 
\be
\delta v_i =\delta v_{0\,i}\;a^{3\,w-1} \, ,
\ee
with  $\delta v_{0\,i}$ an arbitrary  function of $k$. 
From (\ref{1vts}-\ref{2vss}) one can show that all vectors can
be expressed in terms of ${\cal V}_{12}$
that satisfies a second order equation given in 
Appendix \ref{vect}. Thus, only the  vector  ${\cal V}_{12}$ propagates.
For instance, in the case of a radiation dominated universe we
have at the leading order
\bea
\delta v&=& \delta v_0(k)=\text{ constant  in time} \, ;\\[0.2cm]
V_1 &=&-\frac{8}{k^2\,\tau^2}\delta v_0 \, ;\\[0.2cm]
V_2&=& \frac{5}{(k^2\,\tau^2+5)}{\cal V}_{12}'-
\frac{40}{k^2\,\tau^2\,(k^2\,\tau^2+5)}\delta v_0 \, ;\\[0.2cm]
{\cal V}_{12}'' &+&\frac{8 \,k^2\,\tau^2+50}{\tau(k^2\,\tau^2+5)}{\cal
  V}_{12}'
+\frac{3}{\tau^2}(k^2\,\tau^2+5)\,{\cal V}_{12}
-\frac{48 \,k^2\,\tau^2+320}{ \,k^2\,\tau^3\,(k^2\,\tau^2+5)}\delta v_0=0
\, .
\ena
Then for super horizon modes with $k\,\tau\ll 1$ we get
\be
{\cal V}_{12} = C_1 \, 
   \tau^{-\frac{9}{2}-\frac{\sqrt{21}}{2}} + C_2 \, \tau^{\frac{1}{2}
     \left(\sqrt{21}-9\right)}+ \frac{\delta {v_0}}{k^2} \left[\frac{64}{7\, 
   \tau} -\frac{16\, \tau \, k^2}{125} \right]  \, ;
\label{solvect}
\ee
where  $C_{1,2}$   are arbitrary functions of $k$.
For any reasonable choice of  $\delta {v_0}$ there is no growing mode.
The structure of the equations is similar in the case of matter
dominated universe. 

\subsection{Tensor perturbations}
\begin{figure}
\centering
\includegraphics[height=.4\textwidth]{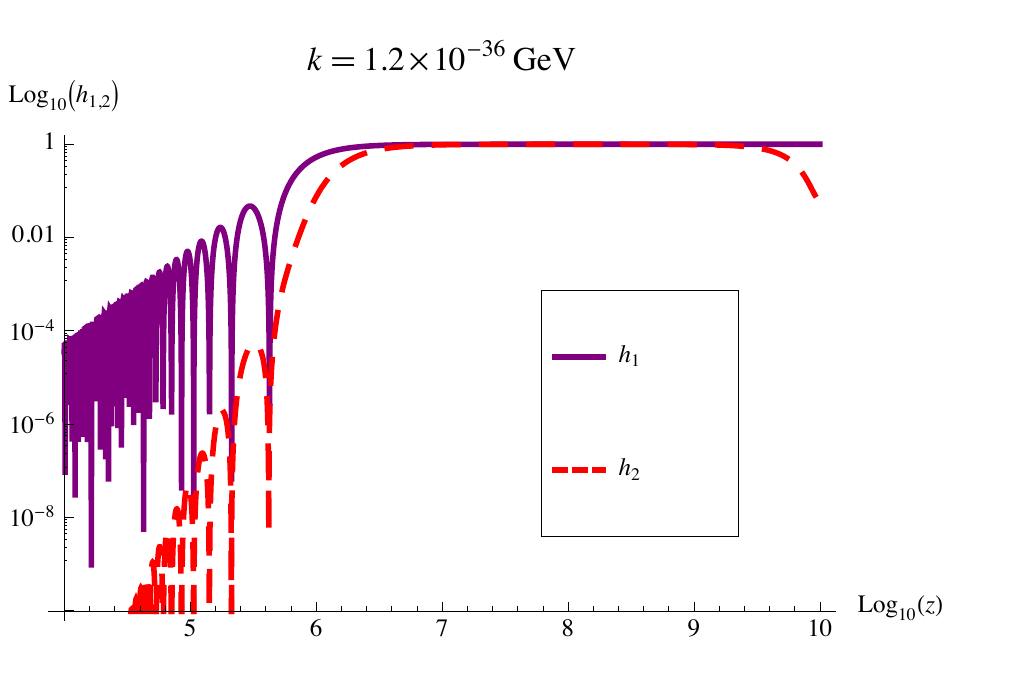}
\caption{Tensor perturbations $h_{1,\,2}$ in a radiation
  dominated universe. 
  The initial conditions are set in deep radiation era ($\Omega_m=0.2$, $H=2\,10^{-42}\,GeV$)  at $z=10^{10}$ as $h_1=1$, $h_2=1/4$, 
  $h_{1,\, 2}'=0$. }
\label{f1}
\end{figure}

The  evolution equations for tensor perturbations  in the radiation era
at next to leading order are
\bea\label{2ten1}
h_1''&+&\frac{2}{\tau}\;h_1'+k^2\;h_1+\epsilon \,\frac{\tau^4}{\tau^4_U}\,\left(\frac{4\,a_0}{5\;\tau}  
h_1'\right)=0 \, ; \\[0.2cm]\label{2ten2}
h_2''&+&\frac{10}{\tau}\;h_2'+25\,k^2\;h_2+\frac{15}{\tau^2} \left(h_1-h_2\right)\;+\\
&& \epsilon \,\frac{\tau^4}{\tau^4_U}\,
\left[-\frac{4 \, \left(a_0 \,\kappa-36\,
a_2\right) }{5\, \kappa\, \tau}\,h_2'-\frac{162\, a_2  }{\kappa\, \tau^2}\,h_1+
  \left(\frac{162 \,a_2 }{\kappa\, \tau^2}-\frac{40 \,k^2 \, \left(a_0
   \,\kappa-6\, a_2\right)}{\kappa  }\right)\, h_2\,\right]=0 \, . \nonumber
\ena 
Tensor  perturbations $h_1$ of  $g$ behave as in  GR, while  the one
of $\tilde g$, $h_2$, beside a sizable coupling with $h_1$ at early times, show  a larger damping factor 
($\frac{2}{\tau}\to\frac{10}{\tau}$) and an  effective  larger mass ($k^2\to 25\,k^2$).
In figure  \ref{f1} we show the numerical solution of (\ref{2ten1}, \ref{2ten2}) .
Clearly there are two regimes, depending on the value of
$k\, \tau$. At very early times, when $k\;\tau \ll 1$, the
two fields $h_{1,2} $ are almost  equal and constant
due to the large coupling proportional to $(h_1-h_2)/ \tau^2 $ present
in (\ref{2ten2}).
As soon as $k\;\tau\sim 1$, $h_1$ starts to oscillate and  triggers the oscillations of the $h_2$ sector (and not the opposite!).
Notice that the damping  of $h_2$ is much larger than the one of $h_1$, indeed
\be
h_1\sim \frac{1}{\tau}, \quad {\rm while}\quad h_2\sim\frac{1}{\tau^5}
\, .
\ee 
During  the matter era, the results are similar.

Summarizing, $h_1$ follows closely GR perturbations; on
superhorizon scales ($k\,\tau\ll1$)  $h_2$ is
proportional to  $h_1$, while  sub-horizon modes are greatly suppressed
($h_2\ll h_1$  when $k\,\tau\gg1$). No instabilities are present.
 
\section{Conclusions}
\label{con}
In this paper we studied cosmological perturbations around FRW background solutions in
a non-linear ghost free massive gravity   theory.
To construct a massive deformation an auxiliary  metric is mandatory.
In the Stuckelberg approach the extra metric is taken to be a
non-dynamical Minkowski metric. Besides  having an ``absolute''
object that take us back to the time of \ae ther-like theories, such a
theories are rather rigid. For instance, there is no black hole solution
featuring modified Newton potential~\cite{ussphe,DeJac} and  there is no
spatially flat FRW solution~\cite{FRWfroz}-\cite{uscosm}, thus a 
spatially negative curvature is required~\cite{uscosm,cur}.
Even if we allow spatial curvature, the resulting solutions feature
an effectively turned off  graviton mass ~\cite{ussphe,uscosm}.
Thus, it is not very surprising that we found the perturbations of 
the branch one  are exactly the same as GR.
The extra modes present in the theory  have zero kinetic term and do
not propagate in a FRW background, this is  typical  of strong
coupling and their propagation is expected at  higher order.
Perturbation theory  around branch one solutions cannot be
trusted due to the presence of the strongly coupled 
extra modes.
In the Stuckelberg approach branch one solutions are the
only available and nothing more can be said, at least using
perturbation theory.

When the second metric is dynamical, the branch two opens up. 
In this case  perturbations  are much richer: all 7+1 expected modes propagate,
and the reliability of perturbation theory results is a non-trivial issue. 
In de Sitter all the extra modes propagate and no instability is 
present  at all the length scales.
In matter or  radiation dominated period, the background solutions is characterized by a
small $\xi$ value proportional to the graviton mass $m$. 
Such a dependence is the origin of deviation from GR in $m\to 0$ limit
with a non trivial $k$ dependence. In particular, the perturbations of the 
metric $g$ that couples  with  matter are very similar to the corresponding ones in GR, while $\tilde g$ perturbations
are not GR-like in the $m\to 0$ limit.
Specifically, we found well behaved  perturbations for
all vector and tensor modes, while one scalar mode shows an
exponential  Jeans-like instability  already well in the radiation
epoch, as soon as its wave length enters inside the
horizon. Thus, though $g$ perturbations are well behaved, such a fast
growth drives sub-horizon scalar perturbations into a non 
perturbative regime just  after a few Hubble times.

A possible way out might be the introduction of a  mirror (dark) matter
sector minimally coupled to the metric $\tilde g$. 
As pointed out in the footnote 3, the small $\xi$ regime is 
the source of instabilities and it can be avoided by the presence of
mirror matter reestablishing the normal power counting for 
small $m $.
The pressure of mirror fluid can contrast the development of sub-horizon instabilities.
Of course, the analysis of such a bi-metric theory plus  matter and mirror
matter is a totally  different ballpark and it requires a dedicated investigation
that will not be given here, we limit to stress that such an
approach would probably not pass the Occam's razor test.

\bigskip

\no {\bf Acknowledgments.} 
We thank F. Nesti for many interesting discussions.
\bigskip

\begin{appendix}
\section{Useful Relations}
\label{rel}
One can verify that the following relations among the additional 
gauge invariant  scalars hold
\bea
&&{\cal B}_2= {\cal B}_1 -(1-c^2) {\cal E}'  \, ;\\[.2cm]
&& {\cal H}_\o \, {\cal F}_2 - {\cal H} \, {\cal F}_1 = ( {\cal H} -
{\cal H}_\o ) (\Phi_1 - \Phi_2) \, ; \\[.2cm]
&& c^2( {\cal F}_2+  {\cal F}_1) =( {\cal B}_1 - {\cal E}) ( {\cal H} +{\cal H}_\o)
-2 c^2  (\Phi_1 - \Phi_2)  \, ;\\[.2cm]
&& c^2( {\cal A}_2-  {\cal A}_1) = ({\cal B}_1 - {\cal E}') \left[ c  (
  {\cal H} -{\cal H}_\o ) -c' \right]  \, ;\\[.2cm]
&& c^2( {\cal A}_2+  {\cal A}_1) =2 c^3 \left(\Psi _2-\Psi _1\right)
+\mathcal{B}_1 \left[ 
   c \left( \mathcal{H}_{\omega }+\mathcal{H}\right) -3 \,  c' \right]+
 3\,  c' \, \mathcal{E}' \\
&&+ c \left[2  \mathcal{B}_1'- 2 \, \mathcal{E}''   -\mathcal{E}'   \left(\mathcal{H}_{\omega
   }+\mathcal{H}\right)\right] \, .
\ena
\section{FRW Background}
\label{corr}

\begin{itemize}
\item During matter era, in the small $\xi $ regime,  we have
$\rho=\rho_0/a^3,\;\;\rho_0\equiv \frac{12}{8 \pi G\,\tau_U^2}$  and for $\tau/\tau_U \ll 1$
\be
a= \frac{\tau ^2}{\tau _U^2} + \frac{a_0 \, 
   \epsilon }{7 }  \frac{\tau ^8}{\tau _U^8}+ \frac{\tau ^{14} }{ \tau
   _U^{14} } \, \frac{3 \, \epsilon^2 \left(4 \, a_0^2 \, \kappa +49
  \,  a_1^2\right)}{637 \, \kappa } + \cdots \; \, .
\ee
\item During radiation era, in the small $\xi $ regime, we have 
$\rho=\frac{\rho_0}{a^4},\;\;\rho_0\equiv\frac{3}{8 \pi G\,\tau_U^2}$ and for $\tau/\tau_U \ll 1$
\be
a=\frac{\tau }{\tau _U}+\epsilon\, 
\frac{a_0 \, \tau ^5}{10\, \tau_U^5}+
\epsilon ^2\,\frac{ \,
   \left(a_0^2\, \kappa+20\,
   a_1^2\right)}{120 \,\kappa\, }\frac{\tau^9 }{\tau _U^9}+...
\ee
\end{itemize}

\section{Scalar Perturbations in the Radiation Era} 
\label{scalrad}

For completeness we give also the leading expressions for the
remaining scalars that can be  expressed in terms of $\Phi_{1/2}$
%(the dotted points stay for corrections of order $\frac{\tau}{\tau_U}\ll 1$)
%
\bea
\Psi_1&=&-\Phi _1 \, ,  \qquad\qquad\qquad\qquad
\Psi_2=-\frac{3 \,\mathcal{E}}{5\, \tau ^2}-\Phi _2 \, ;\\[0.2cm]
u_s &=&\frac{1}{2} \tau  \,\left(\tau \, \Phi _1'+\Phi
   _1\right),\qquad
   \frac{\delta \rho}{\rho}=\frac{2}{3} \,\Phi _1 \left(k^2\, \tau ^2+3\right)+2\, \tau \,\Phi
   _1' \, ;\\[0.2cm]
   {\cal B}_2 &=&-60\,\frac{\mathcal{E}}{\tau}-20\,\tau\,\left(\tau \,  \Phi
   _2'+5 \,\Phi _2\right)  \, ;
%   -\frac{20 \,\left(3 \,\mathcal{E}+\tau ^2 \,\left(\tau \,  \Phi_2'+5 \,\Phi _2\right)\right)}{\tau }+...
   \\[0.2cm]
 {\cal E} &=&  -\frac{\tau ^2\, \left(-9\, \Phi _1 \,\left(k^2\, \tau ^2+9\right)+\Phi _2\,
   \left(2\, k^4 \,\tau ^4+15 \,k^2\, \tau ^2+99\right)+9\, \tau  \,\left(\Phi
   _2'-3\, \Phi _1'\right)\right)}{3\, \left(k^2 \,\tau
   ^2+3\right)^2} \, .
\ena

\section{Scalar Perturbations in the Matter Era} 
\label{scalmatt}
Leading order of the scalar evolution equations during matter era  
\bea\label{matterpert}
\Phi _1'' &+& \frac{6 }{\tau }\Phi _1'%\;\frac{k^2}{3}\;\Phi _1+...
=0 \, ; \\[0.2cm]
\Phi _2'' &+& \frac{4 \left(4 \,x^4+81 \,x^2+720\right)}{\tau\, \left(x
   ^4+18\, x^2+144\right)}\,\Phi _2'+ \frac{-x^6+18\, x ^4+960 \,x^2+11808}{\tau^2 \,\left(x ^4+18 \,x^2+144\right)}\Phi _2-\\ &&
     \frac{6\, \left(x ^4+36\, x ^2+432\right)}{\tau\, \left(x
   ^4+18\, x^2+144\right)}\Phi _1 ' -\frac{x^6+78\, x^4+1728
 \,x^2+14688}{\tau^2 \left(x^4+18\, x ^2+144\right)}  \Phi _1=0 \, .\nonumber
\ena

\section{Next to Leading Corrections for Scalars in  Radiation}
\label{ntl}
For completeness here we give the next to leading correction for the
scalar equations of motion (\ref{scal}). 
\bea
{\cal D}_{1\,1}&=&  4+\epsilon\,\frac{\tau^4}{\tau^4_U}\,\frac{8 a_0    }{45  } +\epsilon^2\,\frac{\tau^8}{\tau^8_U}\,
\frac{4   \,  \left(2\, a_0^2 \,\kappa 
   \left(x^2+3\right)^2+25 \,a_1^2 \,\left(4\, x^4+132\,
   x^2+117\right)\right)}{6075 \,\kappa\,  
   \left(x^2+3\right)^2} \, ;\\[0.2cm]
{\cal D}_{1\,2}&=&  \epsilon ^2\,\frac{\tau^8}{\tau_U^8}\;\frac{2 \,a_1^2\, \left(5\, x^4+6\, x^2+27\right)
   }{9\,\left(x^2+3\right)^2 \kappa } \, ;\\[0.2cm]
{\cal D}_{2\,1}&=&  \frac{36}{x^2+3}+ \epsilon\,\frac{\tau^4}{\tau^4_U}\frac{4  \, \left(3 \,a_0 \,\kappa  \,\left(x^4+9
   x^2-72\right)+10 \,a_2 \,\left(2 x^6+24 x^4+90
   x^2+81\right)\right)}{15\, \kappa\,    \left(x^2+3\right)^3} \, ;\\[0.2cm]
{\cal D}_{2\,2}&=& \frac{10\, x^2+42}{x^2+3}  \nb \\[.2cm]
&-& \epsilon\,\,\frac{\tau^4}{\tau_U^4}\;\frac{4   a_0  
    \left(5 x^6+42 x^4+108 x^2+351\right)+40 \,a_2\, \left(4
    x^6+36 x^4+108 x^2+81\right)}{45\, \kappa  
  \left(x^2+3\right)^3} \, ;\\[0.2cm]
{\cal M}_{1\,1}&=& \frac{x^2}{3}+  \epsilon\,\,\frac{\tau^4}{\tau_U^4}\;\frac{4 \,a_0}{9\, x^2} + 
 \epsilon^2\,\,\frac{\tau^8}{\tau_U^8}\;\frac{  4\, a_0^2\, \kappa\, 
   \left(x^2+3\right)^2+45 \,a_1^2 \,\left(3\, x^4+38
\,   x^2+15\right)}{405 \,\kappa\, x^2\,
   \left(x^2+3\right)^2} \, ;\\[0.2cm]
{\cal M}_{1\,2}&=& \epsilon ^2\,\frac{\tau^8}{\tau_U^8}\; \frac{a_1^2 \left(2 x^6+83 x^4+96
   x^2+459\right)}{9\,\left(x^2+3\right)^2
   \kappa} \, ;
    \\[0.2cm]
{\cal M}_{2\,1}&=& 15+ \frac{72}{x^2+3} \\
&-&  \epsilon\,\,\frac{\tau^4}{\tau_U^4}\,\frac{ \, 36\, a_0 \,\kappa\, \left(x^2-27\right)\,
   \left(x^2+9\right) +10 \,  a_2\, \left(8\, x^8+177 \,x^6+1305 \,x^4+3807 \, x^2+3159\right)}{45\, \kappa  \,
   \left(x^2+3\right)^3}\\[0.2cm]
{\cal M}_{2\,2}&=& \frac{333-5 x^4}{3 x^2+9}  +\epsilon\,\,\frac{\tau^4}{\tau_U^4}\,\left(
\frac{4\, a_0 \,  \left(10 x^8+75
   x^6-54 x^4-1323 x^2-8748\right) 
   }{405   \, \left(x^2+3\right)^3} 
 \right. +\\ && \left.\frac{
 +5\, a_2\, \left(-16 x^8+51
   x^6+1899 x^4+8181 x^2+7533\right)}{405\, \kappa  \,
 \left(x^2+3\right)^3}\right) \, .
\ena 

\section{Vector Perturbations}
\label{vect}
Equation of motion of the vector propagating mode 
\be
\frac{2 \, f_2 \, \kappa  \, k^2 \, \xi ^3 \, \mathcal{H}}{J} \,
\mathcal{V}_{12}'' +\frac{2 \, \kappa
\,    k^2 \, \xi ^2 \, {\cal N}_1 }{J^2} \, \mathcal{V}_{12}'+ f_1 \,
k^2 \, \mathcal{V}_{12} + {\cal N}_0 \, \delta v=0 \, ;
\label{vecteq}
\ee
where
\be
\begin{split}
&{\cal N}_1 =2 a^2 f_2 m^2 \left[2 \mathcal{H}^3 \left(\kappa  \xi^4+\xi ^2\right)+4 \xi  \mathcal{H}^2 \xi '+\xi  \xi '
\mathcal{H}'+\mathcal{H} \left(3 \, \xi '^2-\xi \xi
  ''\right)\right] \\
&+\kappa \,  k^2 \, \xi ^2 \, \left[f_2
   \left(4 \xi  \mathcal{H}^2 \xi '+8 \xi ^2
   \mathcal{H}^3+\xi  \xi ' \mathcal{H}'+\mathcal{H}
   \left(\xi '^2-\xi  \xi ''\right)\right)+\xi 
   \mathcal{H} f_2' \mathcal{H}_{\xi }\right] \, ,
\end{split}
\ee
and
\be
\begin{split}
&J^2 \, {\cal N}_0 = 16 \pi\,  G \,  m^2 \, a^2 \,  \kappa  \, \xi ^2 \,
(w+1) \, \rho _m 
\left \{ 2 a^2 f_2^2 m^2 \left[\xi ' \left(\xi  \left(2
   \mathcal{H}^2+\mathcal{H}'\right)+3 \mathcal{H} \, \xi' \right)-\xi \,
\mathcal{H} \, \xi ''\right] \right. \\
&\left. +\kappa  \, f_2 \,  
   k^2 \, \xi^2 \left[4 \xi ^2 \mathcal{H}^3+2 \xi 
   \mathcal{H}^2 \xi '+\xi  \xi ' \mathcal{H}'+\mathcal{H}
   \left(\xi '^2-\xi  \xi ''\right)\right]+\kappa \, 
    k^2 \, \xi ^3 \, \mathcal{H} \, \mathcal{H}_\xi  
   f_2'\right \} \, .
\end{split}
\ee
All vectors are expressed in terms of ${\cal V}_{12}$, indeed, omitting the spatial
indices, we get
\be
\begin{split}
&\chi = \frac{\kappa\,  k^2\, \xi ^2\, \mathcal{V}_{12}'\,
  \mathcal{H}_{\xi}}{K} -\frac{16 \pi  \,a^2 \,\delta v\, G\, \kappa \, \xi ^2 \,(w+1)
 \,  \mathcal{H}\,\xi  \,\rho }{K} \, , \qquad {\cal H}_\xi = \xi '+2 \,
 \xi \, \mathcal{H} \, ;\\
&K =2 \, m^2 \, a^2 \, f_2 \left[\xi '+\mathcal{H} \left(\kappa \, 
    \xi^3+\xi \right)\right]+\kappa  \, k^2 \, \xi^2 \, \mathcal{H}_{\xi } \, ;
\end{split}
\ee
and
\be
\begin{split}
&V_1 = -\frac{16 \pi  G \,  a^2 \, (w+1) \, \rho\, \delta v \,\left[2 \, m^2\,
a^2 \, f_2 \left(c\,\xi \, \mathcal{H}\right)+\kappa  \, k^2 \, \xi
^2 \, \mathcal{H}_{\xi}\right]}{K \, k^2}-\frac{2\, \kappa\, m^2 \, a^2 \, f_2  \, \xi ^3 \,
 \mathcal{H} \, \mathcal{V}_{12}'}{K} \, , \\
&V_2 =\frac{2 \,m^2  \, a^2 \, f_2 \left(\xi'+\,\xi
 \, \mathcal{H}\right) \, \mathcal{V}_{12}' }{K}-\frac{32 m^2 \, \pi \, G  \, a^4 \,
   f_2 \,  (w+1) \, \rho \, \left(c\,\xi 
   \mathcal{H} \right) \,  \delta v}{K\, k^2} \, .
\end{split}
\ee
\end{appendix}

\end{document}